\documentclass[journal=jacsat,manuscript=article]{achemso}

\usepackage{amsmath}
\usepackage{epstopdf}
\usepackage{graphicx}
\usepackage{dcolumn}
\usepackage{bm}
\usepackage{braket}
\usepackage{gensymb}
\usepackage{array}
\usepackage[utf8]{inputenc}
\usepackage[T1]{fontenc}
\usepackage{mathptmx}

\title{Using the Haken-Strobl-Reineker Model to Determine the Temperature Dependence of the Diffusion Coefficient}

\author{William Barford}
\email{william.barford@chem.ox.ac.uk}
\affiliation{Department of Chemistry, Physical and Theoretical Chemistry Laboratory,\\ University of Oxford, Oxford, OX1 3QZ, United Kingdom}
\begin{document}

\begin{abstract}
Stochastic quantum Liouville equations (SQLE) are widely used to model energy and charge dynamics in molecular systems. The Haken-Strobl-Reineker (HSR) SQLE is a particular paradigm in which the dynamical noise that destroys quantum coherences arises from a white noise (i.e., constant-frequency) spectrum.  A system  subject to the HSR SQLE thus evolves to its `high-temperature' limit, whereby all the eigenstates are equally populated. This result would seem to imply that the predictions of the HSR model, e.g., the temperature dependence of the diffusion coefficient, have no validity for temperatures lower than the particle bandwidth. The purpose of this paper is to show that this assumption is incorrect for translationally invariant systems. In particular, provided that the diffusion coefficient is determined via the mean-squared-displacement, considerations about detailed-balance are irrelevant. Consequently, the high-temperature prediction for the temperature dependence of the diffusion coefficient may be extrapolated to lower temperatures, provided that the bath remains classical.
Thus, for diagonal dynamical disorder the long-time diffusion coefficient, $D_{\infty}(T) = c_{1} /T$, while for both diagonal and off-diagonal disorder, $D_{\infty}(T) =  c_{1}/T + c_{2} T$, where $c_{2} \ll c_{1}$.
An appendix discusses an alternative interpretation from the HSR model of the `quantum to classical' dynamics transition, whereby the dynamics is described as stochastically punctuated coherent motion.
\end{abstract}

\maketitle

%\begin{document}

\section{1. Introduction}\label{Se:1}

Coherent exciton dynamics in static, ordered molecular systems was  described by Merrifield\cite{Merrifield1958} in 1958. Assuming an exciton created  at time $t = 0$ on a monomer, say $n = 0$, he showed that the subsequent wavefunction is $\Psi_n(t) = J_n(2\beta t)$, where $J_n$ is the $n$th order Bessel function of the first-kind and $\beta$ is the intermonomer exciton transfer integral. The wavefunction (illustrated in Fig.\ 2) spreads ballistically with a constant speed and a mean-squared-displacement (MSD) increasing quadratically with time.

As Merrifield observed\cite{Merrifield1958}, however, dynamics on a static, ordered  system is an idealization. Various physical processes, e.g., exciton-phonon coupling, and static and dynamic disorder destroy the coherent motion, eventually causing incoherent (or diffusive) motion where the MSD increases linearly with time. This topic now has a long and rich history, with many reviews  describing the state of the field\cite{Reineker,Capek,May,Dimitriev2022,Barford2022}.

The purpose of this paper is to expand on one particular rich area of investigation, namely the role of thermally-induced noise in destroying quantum coherences. A notable paradigm in this subject is the so-called HSR  stochastic quantum Liouville equation (SQLE), developed and investigated by Haken, Strobl and Reineker\cite{Haken1973,Reineker}. This equation was developed from the underlying time-dependent Schr\"odinger equation (TDSE) assuming that the dynamical fluctuations obey a white-noise spectrum, i.e., a constant power spectrum  (or an Ohmic spectral function). Many important results have been derived from this model\cite{Reineker,Aspuru2009,Knoester2021}. In particular, it describes the `quantum to classical' transition, in which  exciton dynamics exhibits a crossover from ballistic to diffusive as a result of the noise destroying the coherent motion.

As stated, the HSR model assumes a white-noise spectrum, which implies that quantum transitions can occur between any pair of system energy eigenstates. This is turn implies that a system  subject to the HSR SQLE evolves to its `high-temperature' limit, whereby all the eigenstates are equally populated. This result would seem to suggest that the predictions of the HSR model, e.g., the temperature dependence of the diffusion coefficient, have no validity for temperatures lower than the particle bandwidth. The purpose of this paper is to show that this assumption is incorrect for translationally invariant systems. In particular, provided that the diffusion coefficient is determined via the mean-squared-displacement, considerations about detailed-balance are irrelevant. Consequently, the high-temperature prediction for the temperature dependence of the diffusion coefficient may be extrapolated to lower temperatures, provided that the bath remains classical.

This key result will be proved in Section 2. Since the HSR QSLE predictions for the diffusion coefficient in translationally invariant systems are valid for temperatures lower than the particle bandwidth, the predictions of the underlying stochastic TDSE are also equally valid. We use this realization to reinterpret the `quantum to classical' transition as stochastically punctuated coherent motion. This is described in Appendix A.

Unlike the use of the MSD, Appendix B shows that the velocity autocorrelation function cannot be used to extrapolate the HSR predictions to temperatures lower than the bandwidth. Finally, Appendix C derives an expression for the temperature-dependence of the dephasing rate and Appendix D contains some details of the computational techniques.

\section{2. Theory}\label{Se:2}

\subsection{2.1 Model of Exciton Dynamics in Linear Molecular Systems}\label{Se:2.1}

We formulate the problem in terms of Frenkel exciton dynamics in one-dimensional molecular systems, e.g., J-aggregates or conjugated polymers. However, the analysis applies equally to triplet excitons and charges.

The total Hamiltonian is
\begin{equation}\label{}
  \hat{H} = \hat{H}_S + \hat{H}_{SB} + \hat{H}_B,
\end{equation}
where $\hat{H}_S$,  $\hat{H}_{SB}$ and $\hat{H}_B$ are the system, system-bath and bath Hamiltonians, respectively. $\hat{H}_{B}$ is defined in Appendix C (eqn (\ref{Eq:47})).

The system Hamiltonian is defined as
\begin{equation}\label{}
  \hat{H}_S = \alpha \sum_{m=1}^N   |m\rangle \langle m| +  \beta \sum_{m=1}^N \left(|m+1\rangle \langle m| + |m\rangle \langle m+1|\right).
\end{equation}
The ket $|m\rangle$ represents an exciton on monomer $m$, denoting a `site', where $N$ is the number of sites. $\alpha$  and $\beta$ are the onsite potential and  nearest-neighbor exciton transfer integral, respectively.

For a system with translational invariance, the eigenstates of $\hat{H}_S$ are the Bloch states
\begin{equation}\label{Eq:}
|a\rangle = \frac{1}{\sqrt{N}}\sum_{m=1}^N \exp(ik_a m),
\end{equation}
with eigenvalues $E_a = \alpha + 2\beta \cos k_a$,
where $k_a = 2\pi a /N$ is the wavevector. The quantum numbers that label the eigenstates satisfy $1 \le a \le N$.
The particle bandwidth in one-dimension is $4|\beta|$.

$\hat{H}_{SB}$ is the system-bath Hamiltonian
%\begin{widetext}
\begin{equation}\label{}
  \hat{H}_{SB} = \sum_m  \delta\alpha_m(t) |m\rangle \langle m| + \sum_m \delta\beta_m(t)\left(|m+1\rangle \langle m| + |m\rangle \langle m+1| \right),
\end{equation}
%\end{widetext}
where $\delta\alpha_m(t)$ and $\delta\beta_m(t)$ represent dynamical fluctuations in $\alpha$ and $\beta$.
These fluctuations are assumed to be uncorrelated in space, i.e.,
\begin{equation}\label{}
  \langle \delta \alpha_m(t)  \delta \alpha_n(0) \rangle = C_{\alpha}(t) = \sigma_{\alpha}^2 \exp(-t/\tau) \delta_{mn}
\end{equation}
and
\begin{equation}\label{}
  \langle \delta \beta_m(t)  \delta \beta_n(0) \rangle = C_{\beta}(t) = \sigma_{\beta}^2 \exp(-t/\tau) \delta_{mn}.
\end{equation}
In addition, in the white noise limit, defined by $\sigma_{\times} \tau \ll 1$, the bath correlation functions $C_{\times}(t) \rightarrow 2\gamma_{\times} \hbar^2 \delta(t)$, where $\gamma_{\times} = \sigma_{\times}^2 \tau/\hbar^2$ and$\times$ indicates $\alpha$ or $\beta$.
White noise means that the power spectrum, $I(\omega)$, is constant (or the spectral function, $J(\omega) \sim \omega I(\omega) \sim \omega$, i.e., Ohmic), which implies
that the time-dependent part of the Hamiltonian induces transitions between all pairs of eigenstates.

For a classical, harmonic bath, $\gamma_{\times} \propto \sigma_{\times}^2 \propto k_B T $.
More specifically, as shown in Appendix C, for linear system-bath coupling
\begin{equation}\label{Eq:6a}
\gamma_{\times} = \pi k_B T  E_{\times}^r/\hbar^2 \omega_c,
\end{equation}
where $E_{\times}^r$ is the reorganization energy arising from the  system-bath coupling and $\omega_c$ is a high-frequency cut-off for the spectral function.

\subsection{2.2 Determining the Temperature-Dependent Diffusion Coefficient}\label{Se:2.2}

The one-dimensional thermal diffusion coefficient as a function of time is defined as
\begin{equation}\label{Eq:1}
\langle D(t,T) \rangle = \frac{1}{2} \frac{d}{dt} \textrm{Tr}\{ \hat{\rho}(t,T) \hat{x}^2(t)\},
\end{equation}
where $\langle \cdots \rangle$ indicates a thermal average, $\hat{x}$ is the operator for the particle position, and $\hat{\rho}(t,T)$ is the system's reduced density operator.
In the long-time limit  the asymptotic diffusion coefficient is
\begin{equation}\label{Eq:7}
\langle D_{\infty}(T) \rangle  =\frac{1}{2} \frac{d}{dt} \textrm{Tr}\{ \hat{\rho}_0(T) \hat{x}^2(t)\}|_{{\textrm{limit }}t\rightarrow \infty},
\end{equation}
where
\begin{equation}\label{Eq:}
 \hat{\rho}_0(T) = \frac{\exp(-\hat{H}_s/k_B T)}{\textrm{Tr}\{\exp(- \hat{H}_s/k_B T)\}}
\end{equation}
is the equilibrium density operator.

Evaluating the trace over the eigenstate basis of $\hat{H}_S$,  Eqn (\ref{Eq:7}) becomes
\begin{equation}\label{Eq:9}
\langle D_{\infty}(T) \rangle = \sum_a p_a(T) D_a,
\end{equation}
where
\begin{equation}\label{Eq:}
D_a =\frac{1}{2} \frac{d}{dt} \langle a| \hat{x}^2(t)  |a\rangle |_{{\textrm{limit }}t\rightarrow \infty}
\end{equation}
and $p_a(T)$ is the Boltzmann factor.% and $p_a(T) = \exp(-E_a/k_B T)/\sum_a\exp(- E_a/k_B T)$. %and $D_a$ is evaluated for the state $|a\rangle$.

The mean-squared-displacement of a particle prepared in an arbitrary state $|\psi \rangle$ at $t=0$ is defined as
\begin{eqnarray}\label{Eq:11}
\langle \psi | x^2(t) | \psi \rangle = \sum_m \ell^2 m^2 \rho_{mm}(t),
\end{eqnarray}
where $\ell$ is the intermonomer separation and $\rho_{mm}(t)$ are the diagonal elements of the system density matrix in the site basis, $\{|m\rangle\}$. As described in Section 2.3, the density matrix, $\rho_{mn}(t)$, evolves according to an appropriate quantum Liouville equation with the initial condition $\rho_{mn}(0) = \langle m |\psi(0)\rangle\langle \psi(0) |n\rangle$.

Transforming to the eigenstate basis via a transformation matrix, $\textbf{S}$, Eqn (\ref{Eq:11}) becomes
\begin{eqnarray}\label{Eq:12}
\langle \psi | x^2(t) | \psi \rangle = \sum_m \ell^2 m^2 \sum_{a,b} S_{ma} \tilde{\rho}_{ab}(t) S_{bm}^{-1},
\end{eqnarray}
where $\tilde{\rho}_{ab} $
%\begin{equation}\label{Eq:}
%\tilde{\rho}_{ab} = \sum_{m,n} S_{am} \rho_{mn} S_{nb}^{-1}
%\end{equation}
is the density matrix in the eigenstate basis and  $S_{ma} = \langle m | a \rangle$.
For a system with translational invariance, the transformation matrix elements are the Bloch factors
\begin{equation}\label{Eq:13}
S_{ma}  = (S_{am}^{-1})^* = \frac{1}{\sqrt{N}} \exp(-ik_a m).
\end{equation}

Splitting the double sum in Eqn (\ref{Eq:12}) over $a$ and $b$ into the separate sums of $a=b$ and $a \ne b$, and using Eqn (\ref{Eq:13})
we obtain
\begin{eqnarray}\label{Eq:14}
\langle x^2 \rangle &=& \frac{1}{N}\sum_m \ell^2 m^2\left( 1 +  \sum_{a \ne b}  \tilde{\rho}_{ab} \exp(i(k_b-k_a)m) \right),\nonumber\\
\end{eqnarray}
where we have also used $\sum_a  \tilde{\rho}_{aa} = 1$.
%Hence,
%\begin{eqnarray}\label{Eq:}
%\frac{ d\langle x^2 \rangle}{dt} &=& \frac{1}{N}\sum_{m,a \ne b} \ell^2 m^2 \frac{d \tilde{\rho}_{ab}}{dt} \exp(i(k_b-k_a)m).
%\end{eqnarray}
The significance of this result is that it shows that for a translationally invariant system  the diffusion coefficient of a particle prepared in the state $|\psi \rangle$  depends on the evolution of  eigenstate coherences and not  directly on the evolution of  eigenstate populations. Moreover, as will be shown in Section 2.3, for a translationally invariant system, eigenstate coherences  are decoupled from  eigenstate populations, and thus for such systems the diffusion coefficient is completely independent of  eigenstate populations. This means that when evaluating Eqn (\ref{Eq:9}) to determine $\langle D_{\infty}(T)\rangle$ enforcing detailed balance -- or ensuring that  eigenstate populations satisfy their thermal values -- is unnecessary. Thus, $\langle D_{\infty}(T)\rangle$ only depends on temperature \emph{parametrically} via the temperature-dependence of the dephasing rate, Eqn (\ref{Eq:6a}).

Finally, as shown in Section 2.3, the asymptotic diffusion coefficient is independent of the initial state. Denoting this asymptotic value by $D_{\infty}(T)$, setting $D_a = D_{\infty}(T), \forall a$, and using $\sum_a p_a = 1$,  Eqn (\ref{Eq:9}) becomes $\langle D_{\infty}(T) \rangle =  D_{\infty}(T)$.
Thus, our task now is to determine $D_{\infty}(T)$ for a given  dephasing rate. This is achieved via the Haken-Strobl-Reineker model as described in the next section.

%\vfill\pagebreak
\vspace{0.3 cm}

\subsection{2.3 The Haken-Strobl-Reineker Model}\label{Se:2.3}

Haken and Strobl showed that ensemble averages of observables determined via the stochastic TDSE
%, where the Hamiltonian is $\hat{H}_S + \hat{H}_{SB}$,
may be evaluated by a SQLE\cite{Haken1973,Reineker}. For the case of diagonal noise within the site basis, the SQLE reads
%\begin{widetext}
\begin{eqnarray}
 \frac{\partial \rho_{mn}}{\partial t} = - \textrm{i}(\beta/\hbar)\left( \rho_{m\pm 1,n} - \rho_{m,n\pm 1}\right)\delta_{mn}  %\nonumber \\
  -  2 \gamma_{\alpha}\rho_{mn}(1-\delta_{mn}).
\end{eqnarray}
%\end{widetext}
Rotating to the eigenstate basis of $\hat{H}_S$ via Eqn (\ref{Eq:13}), the SQLE becomes for the populations,
\begin{eqnarray}\label{Eq:17a}
 \frac{\partial \tilde{\rho}_{aa}}{\partial t} = -\sum_{b \ne a} \left( k_{ab}\tilde{\rho}_{aa} - k_{ba}\tilde{\rho_{bb}}\right)
\end{eqnarray}
 and for the coherences,
\begin{eqnarray}\label{Eq:17}
 \frac{\partial \tilde{\rho}_{ab}}{\partial t} =
 -   (\textrm{i}\omega_{ab} + 2 \gamma_{\alpha})\tilde{\rho}_{ab} + \frac{2\gamma_{\alpha}}{N}\sum_{c=1}^N \tilde{\rho}_{c,c+(b-a)},
 %(1-\delta_{ab}),
\end{eqnarray}
 where $\omega_{ab} = (E_a - E_b)/\hbar$.

From Eqn (\ref{Eq:17a}) and Eqn (\ref{Eq:17}) we note the following:
\begin{enumerate}
\item{In the HSR model $k_{ab}  = k_{ba} = 2 \gamma_{\alpha}/N$, thus guaranteeing equal eigenstate populations in the long-time limit. In principle, one could impose detailed balance on the rates, thus ensuring that the populations equilibrate to their thermal values. However, as now shown, this is unnecessary.}
\item{The equations of motion for the populations and coherences are decoupled. This proves, by virtue of Eqn (\ref{Eq:14}), that $\langle D_{\infty}(T) \rangle$ is independent of  temperature-dependent populations and only depends on $T$ parametrically via $\gamma_{\alpha}(T)$.}
\item{Eqn (\ref{Eq:17a}) is a single $N$-coupled equation, whereas Eqn (\ref{Eq:17}) are $(N-1) \times N$-coupled equations. Thus, there are $N \times N$-coupled equations in total to solve. A numerical solution is described in Appendix D.}
\end{enumerate}

\section{3. Results}

Haken, Reineker and co-workers derived an expression for $D(t)$ for an initial $\delta$-function source\cite{Schwarzer1972,Reineker}. For diagonal noise this is
\begin{equation}\label{Eq:20}
  D(t) = \frac{\beta^2 \ell^2}{\hbar^2\gamma_{\alpha}}\left( 1 - \exp(-2\gamma_{\alpha}t)\right).
\end{equation}
Reineker\cite{Reineker} also derived  the asymptotic value, $D_{\infty}$, for arbitrary initial conditions. For diagonal noise
\begin{equation}\label{Eq:21}
  D_{\infty} = \beta^2 \ell^2/\hbar^2\gamma_{\alpha}.
\end{equation}

\begin{figure}[h]
\includegraphics[width=0.8\textwidth]{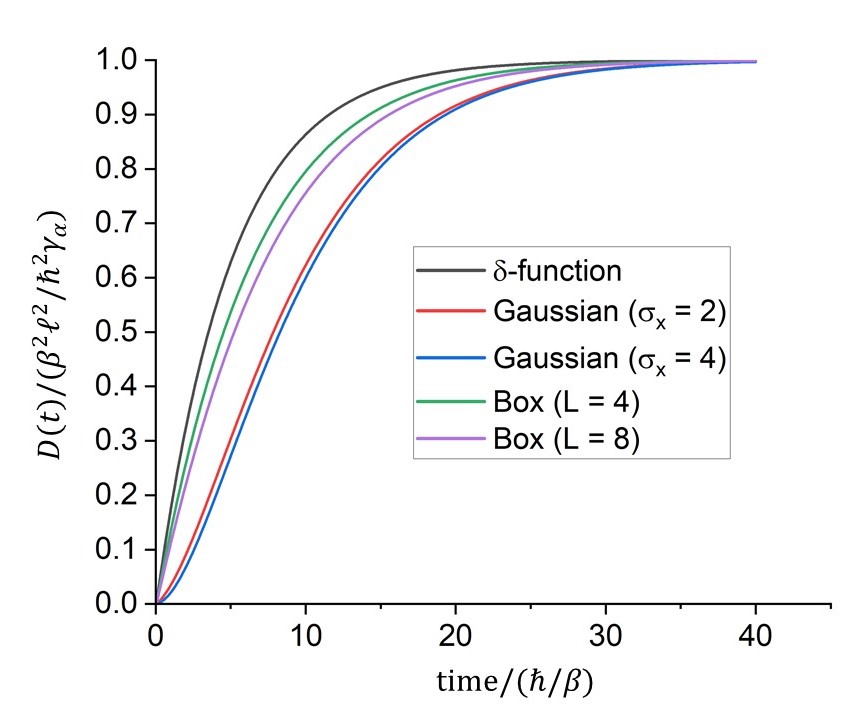}
\label{Fi:1}
\caption{Computed $D(t)$ using Eqn (\ref{Eq:17}) for various initial conditions and diagonal white noise, $\gamma_{\alpha} = 0.1 \beta/\hbar$. For the $\delta$-function source, $D(t)$ satisfies Eqn (\ref{Eq:20}).
The asymptotic values satisfy satisfies Eqn (\ref{Eq:21}). The Gaussian wavefunctions are defined by Eqn (\ref{Eq:A1}). The particle-in-a-box wavefunctions are $\psi_n = (2/(L+1))^{1/2}\sin(\pi n/(L+1))$. The diffusion coefficient at short times for the Gaussian wavepackets is superlinear, as also reported in ref\cite{Cao2023}.}
\end{figure}

These analytical predictions were confirmed by  solving both the HSR SQLE and the stochastic TDSE numerically (as described in Appendix D). Fig.\ 1 shows the numerically evaluated $D(t)$ for different initial conditions, namely a $\delta$-function, Gaussian and particle-in-a-box sources. The $\delta$-function result satisfies Eqn (\ref{Eq:20}), while for all other sources, having a larger initial mean-squared-size and thus a smaller initial mean-squared-speed, the diffusion coefficient increases more slowly with time. However, Eqn (\ref{Eq:21}) is satisfied for all initial conditions.

For translationally invariant systems, the evolution of the populations and coherences are also decoupled in the presence of off-diagonal noise. Thus, for both diagonal and off-diagonal noise, the `high-temperature' limit determined from the HSR SQLE can be extrapolated to temperatures lower than the bandwidth.
Reineker\cite{Reineker} showed that for arbitrary initial conditions\cite{footnote1}
\begin{equation}\label{Eq:22}
 D_{\infty}  =  \frac{\beta^2 \ell^2}{\hbar^2(\gamma_{\alpha}+3 \gamma_{\beta})} + 2\gamma_{\beta}\ell^2,
\end{equation}
where $\gamma_{\alpha}$ and $\gamma_{\beta}$ are the diagonal and off-diagonal dephasing rates, respectively.

Substituting the temperature-dependence of the dephasing rates given by Eqn (\ref{Eq:6a}), we thus have the following prediction for the temperature-dependence of the diffusion coefficient in the presence of white noise:
\begin{equation}\label{Eq:23}
 D_{\infty}(T)  = \frac{c_1}{T} + c_2 T,
\end{equation}
where
\begin{equation}\label{Eq:}
c_1 = \frac{\beta^2 \hbar^2 \omega_c\ell^2}{\pi k_B(E_{\alpha}^r + 3E_{\beta}^r)}
\end{equation}
and
\begin{equation}\label{Eq:}
c_2 = \frac{2\pi k_B E_{\beta}^r\ell^2}{\hbar^2 \omega_c}.
\end{equation}

Equation (\ref{Eq:23}) is valid for translationally invariant systems subject to white noise for all temperatures -- including temperatures lower than the particle bandwidth, $4|\beta|$ -- provided that the bath remains classical. It remains valid if uniform long-range couplings are included, although the coefficients $c_1$ and $c_2$ are altered.

\section{4. Conclusions}\label{Se:4}

This paper has shown that for translationally invariant systems the predictions of the HSR model for the thermal diffusion coefficient can be extrapolated to temperatures lower than the particle bandwidth, provided that the bath remains classical. The proof relies on the observation that for such systems the mean-squared-displacement is independent of  eigenstate populations. Consequently, considerations about detailed balance are irrelevant, and thus the diffusion coefficient  depends on temperature only parametrically via the dephasing rates. When diagonal disorder dominates, $D(T) \sim T^{-1}$. This `high-temperature' limit for $D(T)$ is a common prediction in many theories of charge and energy transport\cite{Fratini}.

Translationally invariant systems subject to white noise are an idealization of more realistic systems, where static disorder,  correlated (non-Markovian) noise and electron-phonon interactions causing polaron formation are all important processes that will modify the predictions of this paper. For such systems, numerically solutions of the TDSE or QLE should explicitly ensure that detailed balance and stationarity are maintained during the system's evolution (e.g., the Redfield quantum Liouville equation of motion\cite{Knoester2016,Nitzan,May,Breuer}, the time-dependent wavepacket diffusion method\cite{Zhong2013,Zhao2015},  hierarchical equations for open systems\cite{Eisfeld2015}, and stochastic Liouville equation methods\cite{Cao2018}). The recently proposed MASH surface-hopping schemes\cite{Mannouch2023a,Runeson2023} are also promising techniques for such simulations.

We conclude by noting that in the presence of static, diagonal disorder the scaling of $D(T)$ with temperature deviates from the HSR prediction\cite{Aspuru2009,Cao2013,Knoester2021}. In particular, for $\gamma \sim T < \beta$, $D(T) \propto T$. This increase of the diffusion coefficient with temperature at low temepatures, sometimes known as environment-assisted quantum transport\cite{Aspuru2009}, is a consequence of the thermal fluctuations destroying the Anderson localization arising from coherent superposition of the particle wavefunction.

\vfill\pagebreak

%\textbf{DATA AVAILABILITY}
%The data that support the findings of this study are available from the corresponding author upon reasonable request.

\vfill\pagebreak

\appendix

\section{Appendix A. The Quantum to Classical Transition}\label{Se:A1}

%The  time-dependent Schr\"odinger equation (TDSE) describes the evolution of a system's wavefunction.
In free space the one-dimensional TDSE reads
\begin{equation}\label{Eq:A1}
  \frac{\partial \Psi(x,t)}{\partial t} =\frac{\textrm{i}\hbar}{2m} \frac{\partial^2 \Psi(x,t)}{\partial x^2},
\end{equation}
where $m$ is the particle mass.
As is well known\cite{Nitzan,Tannor}, a particle prepared in a Gaussian wavepacket at $t=0$, i.e.,
\begin{equation}\label{Eq:A2}
 \Psi(x,0) = \left(\frac{1}{2\pi\sigma_x^2(0)}\right)^{1/4}\exp(-x^2/4\sigma_x^2(0)),
\end{equation}
undergoes a ballistic (or coherent) spread. Thus, its MSD, $\sigma_x^2(t)$, increases as
\begin{equation}\label{}
  \sigma_x^2(t) = \sigma_x^2(0) + \frac{\hbar^2 t^2}{4m\sigma_x^2(0)}.
\end{equation}

In contrast, the diffusion equation describes the evolution of the density of an ensemble of particles, $\rho(x,t)$. In free space the one-dimensional diffusion equation  reads
\begin{equation}\label{}
  \frac{\partial \rho(x,t)}{\partial t} = D \frac{\partial^2 \rho(x,t)}{\partial x^2}.
\end{equation}
Although this equation has precisely the same mathematical form as Eqn (\ref{Eq:A1}), the absence of imaginary i means that its physical predictions are quite different. In particular, an initially localized distribution of particles spreads diffusively, i.e.,
\begin{equation}\label{}
 \rho(x,t) = \left(\frac{1}{2\pi\sigma_x^2(t)}\right)^{1/2}\exp(-x^2/2\sigma_x^2(t)),
\end{equation}
where the MSD increases as
\begin{equation}\label{}
  \sigma_x^2(t) = \sigma_x^2(0) + 2Dt.
\end{equation}

The HSR equation describes how white noise causes a quantum particle's coherent motion to become incoherent. As shown by Reineker\cite{Reineker}, assuming a $\delta$-function source on a lattice, i.e., $|\Psi_{n}(0)|^2 = \delta_{n0}$,  for $t \ll 2\gamma$,
\begin{equation}\label{}
%  \rho_{nn}(t) = J_n^2(2\beta t),
    |\Psi_{n}(t)|^2 = J_n^2(2\beta t),
\end{equation}
(as predicted by the TDSE on a lattice\cite{Merrifield1958}) where $J_n$ is the $n$th-order Bessel function of the first kind.
In contrast, for $t \gg 2\gamma$,
\begin{equation}\label{}
%  \rho_{nn}(t) = \left(\frac{1}{4\pi Dt}\right)^{1/2}\exp(-n^2/4Dt),
|\Psi_{n}(t)|^2 = \left(\frac{1}{4\pi Dt}\right)^{1/2}\exp(-n^2/4Dt),
\end{equation}
(as predicted by the diffusion equation) where $D =  \sigma_x^2(t)/2t = \beta^2/\gamma_{\alpha}$ for diagonal noise.
In general for a $\delta$-function source (where henceforth in this Appendix, we set $\hbar = \ell = 1$)\cite{Reineker,footnote2}
\begin{equation}\label{}
\sigma_x^2(t) = \frac{2\beta^2 t}{\gamma_{\alpha}} + \frac{\beta^2 t}{\gamma_{\alpha}^2}\left(\exp(-2\gamma_{\alpha}t) - 1\right).
\end{equation}

In this interpretation, noise causes a crossover from coherent to incoherent transport. A different interpretation of the `quantum to classical' transition, however, is afforded by the Lindblad formulation of the quantum Liouville equation. As we now show, in this interpretation the particle's dynamics can be viewed as stochastically punctuated coherent motion.

\begin{enumerate}
\item{The Lindblad dissipator is\cite{Breuer}
\begin{equation}\label{}
   \hat{L}[\hat{\rho}(t)] = \Gamma \sum_m \left( \hat{A}_m \hat{\rho}(t) \hat{A}_m^{\dagger} - \frac{1}{2}\left(\hat{A}_m^{\dagger} \hat{A}_m \hat{\rho}(t) + \hat{\rho}(t)  \hat{A}_m^{\dagger} \hat{A}_m\right)\right),
\end{equation}
where $\hat{A}_m$ is the Lindblad jump operator. For diagonal white-noise\cite{Aspuru2009}, $\hat{A}_m = |m\rangle\langle m|$ and $\Gamma = 2\gamma_{\alpha}$.}
\item{An ensemble of quantum trajectories reproduces the observables obtained via a QLE if each trajectory undergoes non-unitary evolution via an effective non-hermitian Hamiltonian\cite{Daley2014}, defined by
\begin{eqnarray}
 \nonumber
  \hat{H}_S \rightarrow \hat{H}_{\textrm{eff}}  &=& \hat{H}_S  - \frac{\textrm{i} \Gamma}{2} \sum_m \hat{A}_m^{\dagger}\hat{A}_m \\
  &=& \hat{H}_S  - \frac{\textrm{i} \Gamma}{2} \sum_m |m\rangle\langle m|.
\end{eqnarray}}
\item{The simulation of a quantum trajectory then proceeds as follows:
\begin{enumerate}
\item{ Given $|\Psi(t)\rangle$ at a time $t$, compute
\begin{equation}\label{}
  |\Psi_{\textrm{trial}}\rangle = \exp(-i \hat{H}_{\textrm{eff}} \delta t) |\Psi(t)\rangle,
\end{equation}
where $\delta t$ is the time step. (See Appendix D for details.)}
\item{Determine $\delta p$, defined via $\langle \Psi_{\textrm{trial}}|\Psi_{\textrm{trial}}\rangle = 1 - \delta p$.}
\item{Then,\\ (i) with a probability $(1-\delta p)$ define
\begin{equation}\label{}
  |\Psi(t+\delta t)\rangle = |\Psi_{\textrm{trial}}\rangle/(1-\delta p)^{1/2}.
\end{equation}
In this case the wavefunction evolves according to $\hat{H}_{\textrm{eff}}$.\\
 Or, \\
(ii) with a probability $\delta p$ define
\begin{eqnarray}\label{Eq:A14}
  |\Psi(t+\delta t)\rangle = \frac{\hat{A}_m |\Psi(t)\rangle}{\langle \Psi(t| \hat{A}_m^{\dagger}\hat{A}_m |\Psi(t)\rangle^{1/2}}.
\end{eqnarray}
Using  the definition that $\hat{A}_m = |m\rangle\langle m|$, Eqn (\ref{Eq:A14}) implies that $|\Psi(t+\delta t)\rangle    = |m\rangle$. Thus, in this case, the wavefunction collapses onto site $m$.
}
\item{If a quantum jump occurs in 3(c)(ii), the site $m$ is chosen with a probability $P_m = |\Psi_m(t)|^2$.}
\end{enumerate}}
\end{enumerate}

Thus, the effect of the Lindblad jump operator in step is 3(c)(ii) to collapse the coherently evolving wavefunction onto the site $m$, after which it then again expands coherently until the next collapse. This process is illustated in Fig.\ 2: At time $t = 0$ the particle is created on site $n_0=0$. Its wavefunction then evolves coherently until the first quantum jump at $t = t_1$. For $t \le t_1$, $\Psi_n(t) = J_n(2\beta t)$, where $J_n$ is the $n$th order Bessel function\cite{Merrifield1958}. At $t=t_1$ the wavefunction collapses onto site $n=n_1$, whereupon it again evolves coherently until a time $t_2$ when it again collapses onto site $n=n_2$.

%\begin{widetext}
\begin{figure}[h]
\includegraphics[width=1.0\textwidth]{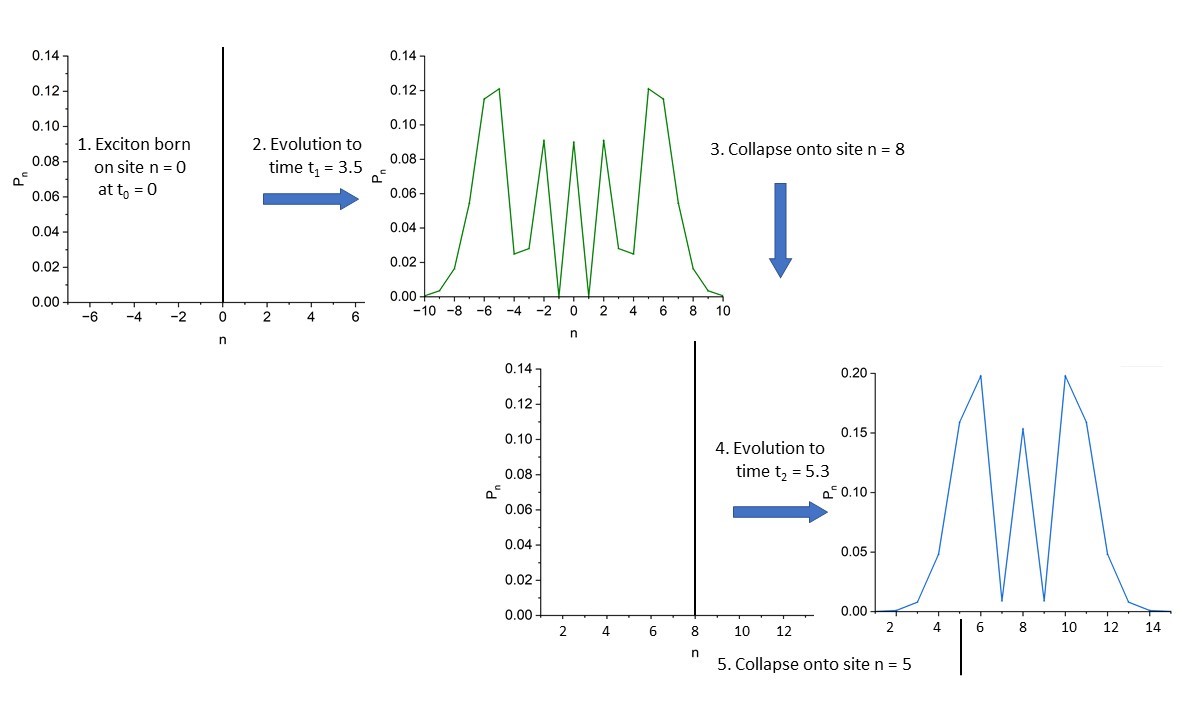}
\caption{An illustration of the evolution of a particle wavefunction subject to the TDSE with stochastic wavefunction collapses caused by white noise. The particle probability, $P_n(t) =  J_n^2(2\beta t)$.
The cumulative effect of the stochastic wavefunction collapses is to cause particle diffusion, where $D  = \beta^2/\gamma_{\alpha}$, in exact agreement with the HSR model. $\gamma_{\alpha} = 0.1 \beta/\hbar$. Time is in units of $\hbar/\beta$.}
\label{Fi:A}
\end{figure}
%\end{widetext}

As these collapses are stochastic in space and time, their cumulative effective is to cause particle diffusion. We can determine the diffusion coefficient as follows.
For a diffusive process the MSD is defined as
\begin{equation}\label{}
  \textrm{MSD}(t) =  \sigma_x^2(t) = N(t) \langle \ell^2 \rangle,
\end{equation}
where  $N(t) = t/\langle \tau \rangle$ is the number of jumps in a time $t$ and  $\langle \tau \rangle = 1/\Gamma =  1/2\gamma_{\alpha}$ is the average time interval between jumps. $\langle \ell^2 \rangle$ is the mean-squared jump size, determined by $\langle \ell^2 \rangle =  \langle v^2\tau^2 \rangle = v^2\langle \tau^2 \rangle$, where $v$ is the particle's coherent speed. For a $\delta$-function source on a lattice, $v = \sqrt{2}\beta$. As determined by simulating the TDSE with Lindblad jump operators, for the  dynamics described here,  $\langle \tau^2 \rangle = 2\langle \tau \rangle^2 = 2/(2\gamma_{\alpha})^2$. Hence,  $\langle \ell^2 \rangle = \beta^2/\gamma_{\alpha}^2$ and therefore $D = N(t) \langle \ell^2 \rangle/2t = \beta^2/\gamma_{\alpha}$, thus -- on average -- rederiving the prediction of the HSR quantum Liouville equation. Ensemble averaging over many particle trajectories will give precisely the same observables (e.g., average particle density) as the stochastic quantum Liouville approach, but the interpretation of the dynamics is different, namely stochastically punctuated coherent dynamics.

%\vfill\pagebreak

\section{Appendix B. Using the Velocity Autocorrelation Function}\label{Se:A2}

\begin{figure}[h]\label{Fi:2}
\includegraphics[width=0.8\linewidth]{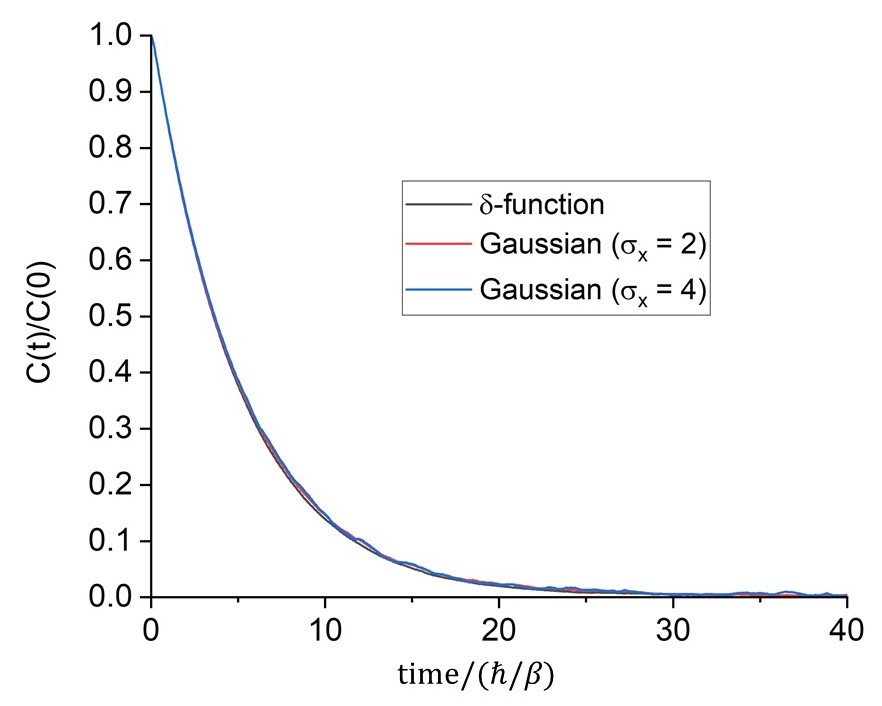}
\caption{The velocity autocorrelation function, $C(t)=\langle  \psi | \hat{C}_v(t) | \psi \rangle$, where $\hat{C}_v(t) = \hat{v}(t)\hat{v}(0)$, computed via the stochastic TDSE with diagonal white noise ($\gamma_{\alpha} = 0.1 \beta/\hbar$).
$C(t)$ satisfies Eqn (\ref{Eq:B4}) for any initial condition.
$C(0) = \langle \hat{v} \rangle_0$ is $2\beta^2\ell^2/\hbar^2$ for a $\delta$-function, while it is $\beta^2/\hbar^2\sigma_x^2$ for a Gaussian wavepacket (defined in Eqn (\ref{Eq:A2})), with $\sigma_x$  in units of $\ell$ and $\hat{v}$ is defined in Eqn (\ref{Eq:57}).
}
\end{figure}

An alternative definition to the thermal diffusion coefficient from Eqn (\ref{Eq:1}) is
\begin{equation}\label{Eq:B1}
\langle D(t,T) \rangle =
%\frac{1}{2}
\int_0^t \textrm{Tr}\{ \hat{\rho}(t',T)  \hat{C}_v(t')\}dt',
\end{equation}
where $\hat{C}_v(t) = \hat{v}(t)\hat{v}(0)$ is the  velocity autocorrelation operator.
%\begin{equation}\label{Eq:}
% \hat{C}_v(t) = \hat{v}(t)\hat{v}(0).
%\end{equation}
The derivation of Eqn (\ref{Eq:B1}) from the identity
\begin{equation}\label{}
  x^2(t) = \int_0^t \int_0^t v(t')v(t'') dt' dt''
\end{equation}
assumes \textit{stationarity}, i.e., that the time-average of $v(t+t')v(t')$ is independent of $t'$.

However, if $\langle \psi |  \hat{C}_v(t) |\psi\rangle$ is evaluated for an arbitrary state $|\psi\rangle$ via the HSR SQLE the stationarity condition is not met. This is because, unlike the case for the mean-squared-displacement, $\langle \psi |  \hat{C}_v(t) |\psi\rangle$ depends on eigenstate populations, which become equally populated under evolution via the HSR SQLE.
%even for translationally invariant systems,

Indeed, as shown in Fig.\ 3,
\begin{equation}\label{Eq:B4}
 C(t)= \langle  \psi | \hat{C}_v(t) | \psi \rangle = \langle \psi|  \hat{C}_v(0) | \psi \rangle \exp(-2\gamma_{\alpha} t),
\end{equation}
%(where $\langle \psi|  \hat{C}_v(0) | \psi \rangle = C(0)$)
in agreement with the definition\cite{Knoester2016,Knoester2021}
\begin{equation}\label{}
  \hat{v}(t) = \exp(\textrm{i}\hat{H}t/\hbar) \hat{v} \exp(-\textrm{i}\hat{H}t/\hbar) \exp(-2\gamma_{\alpha} t)
\end{equation}
for the HSR model. Integrating Eqn (\ref{Eq:B4}) over time gives
\begin{equation}\label{Eq:B5}
  D_{\infty}  = \frac{\langle \psi |\hat{C}_v(0) | \psi\rangle}{2\gamma_{\alpha}},
\end{equation}
which erroneously predicts that the asymptotic  diffusion coefficient depends on the initial condition.

Equation (\ref{Eq:B5}) is only correct for a $\delta$-function source, because this maximally-localized in real-space wavefunction is maximally-delocalized in momentum-space, i.e., all the energy (momentum) eigenstates of a translationally invariant system are already equally populated at $t=0$ so that the stationarity condition is automatically  satisfied. Otherwise, for arbitrary initial conditions, eqn (\ref{Eq:B1}) is only valid provided detailed balance is enforced. Such a scenario is described in ref \cite{Knoester2016}.

%\vfill\pagebreak

\section{Appendix C. The Dephasing Rate}\label{Se:A3}

For linear system-bath coupling, the system-bath Hamiltonian is\cite{Nitzan}
\begin{eqnarray}\label{}
\hat{H}_{SB} = \sum_{m} |m\rangle\langle m| \sum_{j}c_{jm}^{\alpha} u_j + \sum_{m} \left(|m\rangle\langle m+1 | + |m+1\rangle\langle m | \right) \sum_{j}c_{jm}^{\beta} u_j,
\end{eqnarray}
while the bath Hamiltonian of harmonic oscillators is\cite{Nitzan}
\begin{eqnarray}\label{Eq:47}
\hat{H}_{B} = \sum_{j} \left(\dot{u}_j^2 + \omega_j^2 u_j^2\right).
\end{eqnarray}
$u_j$ are the mass-weighted oscillator displacements and
$c_{j}^{\times}$ are the weightings of each normal mode with associated spectral functions,
\begin{equation}\label{}
  J_{\times}(\omega) =  \frac{\pi}{2} \sum_j \frac{(c_j^{\times})^2}{\omega_j} \delta(\omega-\omega_j).
\end{equation}

The bath autocorrelation function associated with diagonal noise is $C_{\alpha}(t) =  \langle \delta \alpha(t)  \delta \alpha(0) \rangle$, where $\delta \alpha(t) = \sum_j c_{j}^{\alpha} u_j(t)$ is the dynamical fluctuation of the on-site potential energy.
For a classical, harmonic bath  and linear system-bath coupling\cite{Nitzan}
\begin{eqnarray}\label{}
 C_{\alpha}(t) = \frac{2k_B T}{\pi} \int_0^{\infty} \textrm{d} \omega\frac{J_{\alpha}(\omega)}{\omega} \cos{\omega t},
\end{eqnarray}
%where $J_{\alpha}(\omega)$ is the bath spectral function for the modes associated with diagonal disorder. 
The spectral function, $J_{\alpha}(\omega)$, for an Ohmic bath with a high-frequency cut-off, $\omega_c$ is defined as
\begin{equation}\label{}
  J_{\alpha}(\omega) =  \left(\frac{\pi E_{\alpha}^r \omega}{\omega_c} \right) \exp(-\omega/\omega_c).
\end{equation}
$E_{\alpha}^r$ is the reorganization energy of the bath modes, satisfying
\begin{equation}\label{}
 E_{\alpha}^r = \frac{1}{\pi} \int_0^{\infty}d\omega \frac{J_{\alpha}(\omega)}{\omega}.
\end{equation}

Thus, using
\begin{eqnarray}\label{}
 \int_0^{\infty}d\omega \exp(-\omega/\omega_c)  \cos{\omega t} & = & \frac{\omega_c}{1 + (\omega_c t)^2} \nonumber \\
              & \rightarrow  & \pi \delta(t),
\end{eqnarray}
as $\omega_c \rightarrow \infty$, we obtain
\begin{eqnarray}\label{Eq:C5}
 C_{\alpha}(t) = \left(\frac{2 \pi k_B T E_{\alpha}^r}{\omega_c}\right) \delta(t).
\end{eqnarray}
Equating Eqn (\ref{Eq:C5}) with the definition of $\gamma_{\alpha}$ given in Section 2.1, i.e., $C_{\alpha}(t) = 2\gamma_{\alpha} \hbar^2 \delta (t)$, yields an expression for the dephasing rate which is linearly proportional to temperature:
\begin{equation}\label{}
  \gamma_{\alpha} = \frac{\pi k_B T  E_{\alpha}^r}{\hbar^2 \omega_c}.
\end{equation}

\section{Appendix D. Computational Methods}\label{Se:A4}

\subsection{D.1 Numerical Solution of the HSR Stochastic Quantum Liouville Equation}

Since the equation of motion for the coherences in the eigenstate basis, Eqn (\ref{Eq:17}), are block diagonal with respect to the quantum number $c$, there are $(N-1) \times N$-coupled equations. These can be cast into the general form
\begin{equation}\label{}
  \frac{d P_{i}}{dt} =  \sum_{j=1}^N L_{ij} P_j,
\end{equation}
for $1 \le i \le (N-1)$, where $\textbf{L}$ is the coupling matrix.
The solution from linear algebra is
\begin{equation}\label{}
  P_i(t) = \sum_{jk} S_{ij} \exp (\lambda_j t) S_{jk}^{-1} P_k(0),
\end{equation}
where $\textbf{S}$ is the matrix whose columns are the eigenvectors of $\textbf{L}$, $\{ \lambda \}$ are the corresponding eigenvalues and $P_k(0)$ is an initial condition. To produce the results illustrated in Fig.\ 1, the complex matrix $\textbf{L}$ was diagonalized via the LAPACK routine  \textsf{ZGEEV}, and the complex matrix $\textbf{S}$ was inverted via the LAPACK routines  \textsf{ZGETRF} and \textsf{ZGETRI}.

\subsection{D.2 Numerical Solution of the Time-Dependent Schrodinger Equation}

In general, given $\Psi(t)$ we require
\begin{equation}\label{}
  \Psi(t+\delta t)  = \exp(-\textrm{i} (H_S + H_{SB}(t))\delta t/\hbar) \Psi(t).
\end{equation}
This is easily accomplished via the \textit{Short Iterative Lanczos Propagator} method\cite{Tannor}, whereby a Krylov space of $N$ vectors is generated via the Lanczos method. Diagonalizing the tridiagonal Hamiltonian, $\textbf{H}_{\textrm{Lanczos}}$, within the Krylov space yields
\begin{equation}\label{}
  \Psi(t+\delta t)  = \textbf{S}\cdot \exp(-\textrm{i} \textbf{D}\delta t/\hbar)\cdot \textbf{S}^{\dagger}\Psi(t),
\end{equation}
where $\textbf{S}$ is the is the matrix whose columns are the eigenvectors of $\textbf{H}_{\textrm{Lanczos}}$, the diagonal elements of $\textbf{D}$ are its eigenvalues, and $\Psi(t)$ is the first vector in the Krylov space.

The expectation value of the velocity autocorrelation function is
\begin{eqnarray}
\langle \Psi | \hat{C}_v(t) |\Psi\rangle &=& \langle \Psi | \hat{v}(t)\hat{v}(0) |\Psi\rangle \nonumber \\
    &=& \langle \Psi (t) | \hat{v} |\Phi(t)\rangle,
\end{eqnarray}
where  $|\Phi\rangle  = \hat{v} | \Psi\rangle$
and the velocity operator on a lattice  is
\begin{equation}\label{Eq:57}
 \hat{v} = \textrm{i} (\ell \beta/\hbar) \sum_m \left( |m\rangle\langle m + 1 | - |m+1 \rangle\langle m  | \right).
\end{equation}

\vfill\pagebreak

%\bibliography{references}

\begin{mcitethebibliography}{27}
\providecommand*\natexlab[1]{#1}
\providecommand*\mciteSetBstSublistMode[1]{}
\providecommand*\mciteSetBstMaxWidthForm[2]{}
\providecommand*\mciteBstWouldAddEndPuncttrue
  {\def\EndOfBibitem{\unskip.}}
\providecommand*\mciteBstWouldAddEndPunctfalse
  {\let\EndOfBibitem\relax}
\providecommand*\mciteSetBstMidEndSepPunct[3]{}
\providecommand*\mciteSetBstSublistLabelBeginEnd[3]{}
\providecommand*\EndOfBibitem{}
\mciteSetBstSublistMode{f}
\mciteSetBstMaxWidthForm{subitem}{(\alph{mcitesubitemcount})}
\mciteSetBstSublistLabelBeginEnd
  {\mcitemaxwidthsubitemform\space}
  {\relax}
  {\relax}

\bibitem[Merrifield(1958)]{Merrifield1958}
Merrifield,~R.~E. Propagation of Electronic Excitation in Insulating Crystals.
  \emph{Journal of Chemical Physics} \textbf{1958}, \emph{28}, 647--650\relax
\mciteBstWouldAddEndPuncttrue
\mciteSetBstMidEndSepPunct{\mcitedefaultmidpunct}
{\mcitedefaultendpunct}{\mcitedefaultseppunct}\relax
\EndOfBibitem
\bibitem[Kenkre and Reineker(1982)Kenkre, and Reineker]{Reineker}
Kenkre,~V.~M.; Reineker,~P. \emph{Exciton dynamics in molecular crystals and
  aggregates}; Springer-Verlag: Berlin; New York, 1982\relax
\mciteBstWouldAddEndPuncttrue
\mciteSetBstMidEndSepPunct{\mcitedefaultmidpunct}
{\mcitedefaultendpunct}{\mcitedefaultseppunct}\relax
\EndOfBibitem
\bibitem[Silinish and Capek(1994)Silinish, and Capek]{Capek}
Silinish,~E.; Capek,~V. \emph{Exciton dynamics in molecular crystals and
  aggregates}; American Institute of Physics: New York, 1994\relax
\mciteBstWouldAddEndPuncttrue
\mciteSetBstMidEndSepPunct{\mcitedefaultmidpunct}
{\mcitedefaultendpunct}{\mcitedefaultseppunct}\relax
\EndOfBibitem
\bibitem[May and K\"uhn(2011)May, and K\"uhn]{May}
May,~V.; K\"uhn,~O. \emph{Charge and energy transfer dynamics in molecular
  systems}; Wiley-VCH: Weinheim, 2011\relax
\mciteBstWouldAddEndPuncttrue
\mciteSetBstMidEndSepPunct{\mcitedefaultmidpunct}
{\mcitedefaultendpunct}{\mcitedefaultseppunct}\relax
\EndOfBibitem
\bibitem[Dimitriev(2022)]{Dimitriev2022}
Dimitriev,~O.~P. Dynamics of Excitons in Conjugated Molecules and Organic
  Semiconductor Systems. \emph{Chemical Reviews} \textbf{2022}, \emph{122},
  8487--8593\relax
\mciteBstWouldAddEndPuncttrue
\mciteSetBstMidEndSepPunct{\mcitedefaultmidpunct}
{\mcitedefaultendpunct}{\mcitedefaultseppunct}\relax
\EndOfBibitem
\bibitem[Barford(2022)]{Barford2022}
Barford,~W. Exciton dynamics in conjugated polymer systems. \emph{Frontiers in
  Physics} \textbf{2022}, \emph{10:1004042}\relax
\mciteBstWouldAddEndPuncttrue
\mciteSetBstMidEndSepPunct{\mcitedefaultmidpunct}
{\mcitedefaultendpunct}{\mcitedefaultseppunct}\relax
\EndOfBibitem
\bibitem[Haken and Strobl(1973)Haken, and Strobl]{Haken1973}
Haken,~H.; Strobl,~G. Exactly Solvable Model for Coherent and Incoherent
  Exciton Motion. \emph{Zeitschrift Fur Physik} \textbf{1973}, \emph{262}\relax
\mciteBstWouldAddEndPuncttrue
\mciteSetBstMidEndSepPunct{\mcitedefaultmidpunct}
{\mcitedefaultendpunct}{\mcitedefaultseppunct}\relax
\EndOfBibitem
\bibitem[Rebentrost \latin{et~al.}(2009)Rebentrost, Mohseni, Kassal, Lloyd, and
  Aspuru-Guzik]{Aspuru2009}
Rebentrost,~P.; Mohseni,~M.; Kassal,~I.; Lloyd,~S.; Aspuru-Guzik,~A.
  Environment-assisted quantum transport. \emph{New Journal of Physics}
  \textbf{2009}, \emph{11}, 033003\relax
\mciteBstWouldAddEndPuncttrue
\mciteSetBstMidEndSepPunct{\mcitedefaultmidpunct}
{\mcitedefaultendpunct}{\mcitedefaultseppunct}\relax
\EndOfBibitem
\bibitem[Kunsel \latin{et~al.}(2021)Kunsel, Jansen, and Knoester]{Knoester2021}
Kunsel,~T.; Jansen,~T. L.~C.; Knoester,~J. Scaling relations of exciton
  diffusion in linear aggregates with static and dynamic disorder.
  \emph{Journal of Chemical Physics} \textbf{2021}, \emph{155}, 134305\relax
\mciteBstWouldAddEndPuncttrue
\mciteSetBstMidEndSepPunct{\mcitedefaultmidpunct}
{\mcitedefaultendpunct}{\mcitedefaultseppunct}\relax
\EndOfBibitem
\bibitem[Schwarzer and Haken(1972)Schwarzer, and Haken]{Schwarzer1972}
Schwarzer,~E.; Haken,~H. Moments of Coupled Coherent and Incoherent Motion of
  Excitons. \emph{Physics Letters A} \textbf{1972}, \emph{A 42}, 317--318\relax
\mciteBstWouldAddEndPuncttrue
\mciteSetBstMidEndSepPunct{\mcitedefaultmidpunct}
{\mcitedefaultendpunct}{\mcitedefaultseppunct}\relax
\EndOfBibitem
\bibitem[Tutunnikov \latin{et~al.}(2023)Tutunnikov, Chuang, and Cao]{Cao2023}
Tutunnikov,~I.; Chuang,~C.~R.; Cao,~J.~S. Coherent Spatial Control of Wave
  Packet Dynamics on Quantum Lattices. \emph{Journal of Physical Chemistry
  Letters} \textbf{2023}, \emph{14}, 11632--11639\relax
\mciteBstWouldAddEndPuncttrue
\mciteSetBstMidEndSepPunct{\mcitedefaultmidpunct}
{\mcitedefaultendpunct}{\mcitedefaultseppunct}\relax
\EndOfBibitem
\bibitem[foo()]{footnote1}
Equation (\ref{Eq:22}) was confirmed by numerical solutions of the stochastic
  TDSE. \relax
\mciteBstWouldAddEndPunctfalse
\mciteSetBstMidEndSepPunct{\mcitedefaultmidpunct}
{}{\mcitedefaultseppunct}\relax
\EndOfBibitem
\bibitem[Fratini \latin{et~al.}(2016)Fratini, Mayou, and Ciuchi]{Fratini}
Fratini,~S.; Mayou,~D.; Ciuchi,~S. The Transient Localization Scenario for
  Charge Transport in Crystalline Organic Materials. \emph{Advanced Functional
  Materials} \textbf{2016}, \emph{26}, 2292--2315\relax
\mciteBstWouldAddEndPuncttrue
\mciteSetBstMidEndSepPunct{\mcitedefaultmidpunct}
{\mcitedefaultendpunct}{\mcitedefaultseppunct}\relax
\EndOfBibitem
\bibitem[Chuang \latin{et~al.}(2016)Chuang, Lee, Moix, Knoester, and
  Cao]{Knoester2016}
Chuang,~C.; Lee,~C.~K.; Moix,~J.~M.; Knoester,~J.; Cao,~J.~S. Quantum Diffusion
  on Molecular Tubes: Universal Scaling of the 1D to 2D Transition.
  \emph{Physical Review Letters} \textbf{2016}, \emph{116}, 196803\relax
\mciteBstWouldAddEndPuncttrue
\mciteSetBstMidEndSepPunct{\mcitedefaultmidpunct}
{\mcitedefaultendpunct}{\mcitedefaultseppunct}\relax
\EndOfBibitem
\bibitem[Nitzan(2006)]{Nitzan}
Nitzan,~A. \emph{Chemical dynamics in condensed phases: relaxation, transfer
  and reactions in condensed molecular systems}; Oxford University Press:
  Oxford; New York, 2006\relax
\mciteBstWouldAddEndPuncttrue
\mciteSetBstMidEndSepPunct{\mcitedefaultmidpunct}
{\mcitedefaultendpunct}{\mcitedefaultseppunct}\relax
\EndOfBibitem
\bibitem[Breuer and Petruccione(2002)Breuer, and Petruccione]{Breuer}
Breuer,~H.-P.; Petruccione,~F. \emph{The theory of open quantum systems};
  Oxford University Press: Oxford; New York, 2002\relax
\mciteBstWouldAddEndPuncttrue
\mciteSetBstMidEndSepPunct{\mcitedefaultmidpunct}
{\mcitedefaultendpunct}{\mcitedefaultseppunct}\relax
\EndOfBibitem
\bibitem[Zhong and Zhao(2013)Zhong, and Zhao]{Zhong2013}
Zhong,~X.~X.; Zhao,~Y. Non-Markovian stochastic Schrodinger equation at finite
  temperatures for charge carrier dynamics in organic crystals. \emph{Journal
  of Chemical Physics} \textbf{2013}, \emph{138}, 014111\relax
\mciteBstWouldAddEndPuncttrue
\mciteSetBstMidEndSepPunct{\mcitedefaultmidpunct}
{\mcitedefaultendpunct}{\mcitedefaultseppunct}\relax
\EndOfBibitem
\bibitem[Han \latin{et~al.}(2015)Han, Ke, Zhong, and Zhao]{Zhao2015}
Han,~L.; Ke,~Y.~L.; Zhong,~X.~X.; Zhao,~Y. Time-Dependent Wavepacket Diffusion
  Method and its Applications in Organic Semiconductors. \emph{International
  Journal of Quantum Chemistry} \textbf{2015}, \emph{115}, 578--588\relax
\mciteBstWouldAddEndPuncttrue
\mciteSetBstMidEndSepPunct{\mcitedefaultmidpunct}
{\mcitedefaultendpunct}{\mcitedefaultseppunct}\relax
\EndOfBibitem
\bibitem[Suess \latin{et~al.}(2015)Suess, Strunz, and Eisfeld]{Eisfeld2015}
Suess,~D.; Strunz,~W.~T.; Eisfeld,~A. Hierarchical Equations for Open System
  Dynamics in Fermionic and Bosonic Environments. \emph{Journal of Statistical
  Physics} \textbf{2015}, \emph{159}, 1408--1423\relax
\mciteBstWouldAddEndPuncttrue
\mciteSetBstMidEndSepPunct{\mcitedefaultmidpunct}
{\mcitedefaultendpunct}{\mcitedefaultseppunct}\relax
\EndOfBibitem
\bibitem[Hsieh and Cao(2018)Hsieh, and Cao]{Cao2018}
Hsieh,~C.~Y.; Cao,~J.~S. A unified stochastic formulation of dissipative
  quantum dynamics. I. Generalized hierarchical equations. \emph{Journal of
  Chemical Physics} \textbf{2018}, \emph{148}, 014103\relax
\mciteBstWouldAddEndPuncttrue
\mciteSetBstMidEndSepPunct{\mcitedefaultmidpunct}
{\mcitedefaultendpunct}{\mcitedefaultseppunct}\relax
\EndOfBibitem
\bibitem[Mannouch and Richardson(2023)Mannouch, and Richardson]{Mannouch2023a}
Mannouch,~J.~R.; Richardson,~J.~O. A mapping approach to surface hopping.
  \emph{Journal of Chemical Physics} \textbf{2023}, \emph{158}, 104111\relax
\mciteBstWouldAddEndPuncttrue
\mciteSetBstMidEndSepPunct{\mcitedefaultmidpunct}
{\mcitedefaultendpunct}{\mcitedefaultseppunct}\relax
\EndOfBibitem
\bibitem[Runeson and Manolopoulos(2023)Runeson, and Manolopoulos]{Runeson2023}
Runeson,~J.~E.; Manolopoulos,~D.~E. A multi-state mapping approach to surface
  hopping. \emph{Journal of Chemical Physics} \textbf{2023}, \emph{159},
  094115\relax
\mciteBstWouldAddEndPuncttrue
\mciteSetBstMidEndSepPunct{\mcitedefaultmidpunct}
{\mcitedefaultendpunct}{\mcitedefaultseppunct}\relax
\EndOfBibitem
\bibitem[Moix \latin{et~al.}(2013)Moix, Khasin, and Cao]{Cao2013}
Moix,~J.~M.; Khasin,~M.; Cao,~J.~S. Coherent quantum transport in disordered
  systems: I. The influence of dephasing on the transport properties and
  absorption spectra on one-dimensional systems. \emph{New Journal of Physics}
  \textbf{2013}, \emph{15}, 085010\relax
\mciteBstWouldAddEndPuncttrue
\mciteSetBstMidEndSepPunct{\mcitedefaultmidpunct}
{\mcitedefaultendpunct}{\mcitedefaultseppunct}\relax
\EndOfBibitem
\bibitem[Tannor(2007)]{Tannor}
Tannor,~D.~J. \emph{Introduction to quantum mechanics: a time-dependent
  perspective}; University Science Books: Sausalito, Calif., 2007\relax
\mciteBstWouldAddEndPuncttrue
\mciteSetBstMidEndSepPunct{\mcitedefaultmidpunct}
{\mcitedefaultendpunct}{\mcitedefaultseppunct}\relax
\EndOfBibitem
\bibitem[foo()]{footnote2}
Likewise, the quantum to classical transition for an initial Gaussian
  wavefunction may be inferred by time-integrating the diffusion coefficient
  for the Gaussian initial state, as shown in Fig.\ 1, to obtain MSD(t). \relax
\mciteBstWouldAddEndPunctfalse
\mciteSetBstMidEndSepPunct{\mcitedefaultmidpunct}
{}{\mcitedefaultseppunct}\relax
\EndOfBibitem
\bibitem[Daley(2014)]{Daley2014}
Daley,~A.~J. Quantum trajectories and open many-body quantum systems.
  \emph{Advances in Physics} \textbf{2014}, \emph{63}, 77--149\relax
\mciteBstWouldAddEndPuncttrue
\mciteSetBstMidEndSepPunct{\mcitedefaultmidpunct}
{\mcitedefaultendpunct}{\mcitedefaultseppunct}\relax
\EndOfBibitem
\end{mcitethebibliography}

\providecommand{\latin}[1]{#1}
\makeatletter
\providecommand{\doi}
  {\begingroup\let\do\@makeother\dospecials
  \catcode`\{=1 \catcode`\}=2 \doi@aux}
\providecommand{\doi@aux}[1]{\endgroup\texttt{#1}}
\makeatother
\providecommand*\mcitethebibliography{\thebibliography}
\csname @ifundefined\endcsname{endmcitethebibliography}
  {\let\endmcitethebibliography\endthebibliography}{}

\end{document}